\documentstyle[12pt]{article}
\begin{document}
\begin{center}{\Large {\bf Random field Ising model swept by propagating
magnetic field wave:
Athermal nonequilibrium phase diagram}}\end{center}
\vskip 1cm
\begin{center}
{\it Muktish Acharyya}\\
{\it Department of Physics, Presidency University}\\
{\it 86/1 College Street, Calcutta-700073, India}\\
{E-mail:muktish.physics@presiuniv.ac.in}\end{center}

\vskip 2cm

The dynamical steady state behaviour of the random field
 Ising ferromagnet swept
by a propagating magnetic field wave is studied at zero temperature by Monte
Carlo simulation in two dimensions. The distribution of the random
 field is bimodal type. For
a fixed set of values of the frequency, wavelength and amplitude
of propagating magnetic 
field wave and the strength of the random field, four distinct 
dynamical steady states or nonequilibrium phases were identified. 
These four nonequilibrium
phases are characterised by different values of structure factors.
State or phase of first kind, where all spins are parallel (up). 
This phase is a frozen or pinned where the propagating field has no effect.
The second one is,
the
propagating type, where the sharp strips formed by parallel spins 
are found to move coherently. The third 
one is also propagating type,  
where the boundary of the strips of spins is not very sharp. The fourth kind, shows no propagation
of strips of magnetic spins, forming a 
homogeneous distribution of up and down spins. This is disordered phase.
The existence of these four
dynamical phases or modes depends on the value of the amplitude of propagating 
magnetic field wave and the strength of random (static) field. A phase diagram 
has also been drawn, in the plane formed by the amplitude of propagating field and the
strength of random field. It is also checked that, the existence of these dynamical
phases is neither a finite size effect nor a transient phenomenon.



\newpage

\noindent {\bf I. Introduction:}

The behaviour of
Ising ferromagnet with randomly quenched magnetic field is an active
field of modern research. The zero temperature ferro-para transition
\cite{imry}
by the random field, made it more interesting. A simple model was proposed
to understand the hysteresis incorporating the return point memory
\cite{shore} and Barkhausen noise\cite{brien}. Later, the statistics
\cite{tadic1} and
the dynamic critical behaviour\cite{tadic2} of 
Barkhausen avalanches were studied
in random field Ising model (RFIM). The hysteresis loop was exactly
determined in one dimensional RFIM\cite{sukla}. The RFIM was also studied
\cite{dhar} in the Bethe lattice. 

The dynamics of the domain wall in RFIM is another area
 of interest in 
modern research of statistical physics. The motion of domain wall shows
very interesting depinning transition at zero temperature. Due to the
energy barriers created by disorder (random field), the domain wall is 
pinned and the motion of the domain wall remains stopped upto a critical
field\cite{nowak, kolton}. However, at any finite temperature the
depinning transition is softened and the thermal fluctuation assists to
overcome the energy barrier. Thus with the application
of sufficiently small amount of field, the 
motion of domain
wall can reach a nonzero mean velocity, which is known as creep motion
\cite{lemerele,lubeck,rosso}. Very recently, the creep motion of domain wall
in the two dimensional RFIM was studied (by Monte Carlo simulations) with a
driving field\cite{epl12} and observed field-velocity relationship and
estimated the creep exponent.

All the above mentioned studies are done with constant and uniform magnetic
field applied to RFIM. The behaviour of RFIM in the presence of time dependent
magnetic field is not yet studied widely. A study, with time dependent
(but uniform over the space) magnetic field applied to RFIM, shows dynamical
symmetry breaking and athermal nonequilibrium phase transition\cite{marfim}. The 
dynamical phase boundary was also drawn in the plane formed by the width 
of the disorder 
(random field) and the amplitude of the oscillating 
magnetic field. This study with
time dependent magnetic field, showing dynamic symmetry breaking transition,
is unable to extract any information about the morphology of the system.
It would be interesting, to study the dynamic {\it structural}
behaviours of RFIM, one has to use propagating magnetic field. In this paper,
the dynamic behaviours of spins of RFIM swept by propagating magnetic
field are studied by Monte Carlo simulation. The paper is organised as follows:
the model and the simulation technique is discussed in section II, 
the numerical or simulational results are given in 
section III and the paper ends with
a summary in section IV.

\vskip 1cm

\noindent {\bf II. The model and simulation:}

The time dependent Hamiltonian representing the random field Ising model 
swept by a propagating
magnetic field wave may be written as
\begin{equation}
H(t) = -J\Sigma s(x,y,t) s(x',y',t) 
-\Sigma h(x,y) s(x,y,t) -\Sigma hp(y,t) s(x,y,t)
\end{equation}
\noindent The $s(x,y,t)$ represents the 
Ising spin variable ($\pm 1$) at lattice
site $(x,y)$ at time $t$ on a square lattice of linear size $L$.
$J (> 0)$ is the ferromagnetic (uniform) interaction strength. The first sum
for Ising spin-spin interaction is carried over the nearest neighbours only.
$h(x,y)$ is the random field acting at ($x,y$) lattice site. 
The distribution of the
random field is bimodal type, and represented as
\begin{equation}
P(h(x,y)) = {1 \over 2}\delta (h(x,y)-w)+ {1 \over 2}\delta (h(x,y)+w)
\end{equation}
Here, $w$ represents the width of random field.
The $hp(y,t)$ is the value of the propagating (along the positive y-direction)
 magnetic field at time $t$ acting at ($x,y$) lattice site.
\begin{equation}
hp(y, t) = h_0 {\rm cos} (2\pi f t -2\pi y/\lambda)
\end{equation}
The $h_0$, $f$ and $\lambda$ represent the amplitude, frequency and the
wavelength respectively of the propagating magnetic field wave.
We have considered the initial ($t=0$) configuration as all spins are up 
($s(x,y,t=0) = +1$ for all $x$ and $y$)). The spins are updated
randomly (a site ($x,y$) is chosen at random) and spin flip occurs 
only when the energy decreases
due to flipping (the zero temperature
Metropolis algorithm). The boundary condition is periodic
in both ($x$ and $y$) direction. Here, the value of magnetic field is
measured in the unit of $J$.

\vskip 1cm

\noindent {\bf III. The Results:}

 The simulation was performed on a square lattice of linear size $L=100$ with
periodic boundary condition. The lattice spacing is taken as unity. Initially
all the spin variables assume the value $+1$. A lattice site 
($x,y$) is chosen randomly
and the spin variable is updated/flipped only if the flipping reduces
the energy. $L^2$ such random updates of spin variables is defined as
unit Monte Carlo Step. The time $t$, in the problem is measured
in this unit. 

The frequency ($f$) and the wavelength ($\lambda$)
of the propagating magnetic field wave are kept fixed ($f=0.01$ and 
$\lambda=20$) throughout the present study. The long time behaviour
and the dynamical steady states are studied here. 
 As the wave propagtes through the system of random field Ising model, the
dynamical steady states are observed to form in vareity of morphology depending
upon the values of $h_0$ and $w$. One such vareity of morphology 
(the spin configurations on the lattice) are depicted
in Figure-1. Here, figure-1a, shows the strong ferromagnetic (ordered) phase (where
all spins are up) for $h_0=3.5$ and $w=0.4$. 
This (phase-1) is a non-propagating frozen or pinned phase.
As the width ($w$)
of random field increases to $w=1.0$, the propagating dynamical mode is 
observed. Here, the groups of parallel spins forms strips and these strips
of alternate down and up spins
propagates along the direction of propagation of the magnetic field wave.
The wavelength of this propagation is same as that of 
externally applied propagating magnetic field wave.
This coherent propagation of strips of alternate
down and up spins is shown in figure-1b. 
Here, the groups of down spins forms strips of sharp
boundary. The strips formed by parallel spins are very compact in structure.
No down spin can be observed in the strip formed by up spins. These strips
propagates along the direction of propagation of the magnetic field wave.
This coherent propagation of strips of
down spins forms phase-2.
After further increases of the width ($w=4.0$)
of random field, another dynamical phase appears. In this phase, the 
strips of down spins are formed but the boundary of these strips are not
as sharp as that observed in phase-2. 
In this phase, the strips are not very compact. The down spin may be found in the
strip formed by up spins or vice versa.
Here also, the propagation
of these noncompact strips is observed. This is phase-3 and shown in
figure-1c. The fourth kind of dynamical phase, i.e., phase-4 appears for
larger values of width ($w=6.0$) of the random field. 
This is almost a homogeneous
mixtures of up and down spins and constitutes phase-4 or disordered phase.
This is non-propagating in nature and shown in figure-1d for $w=6.0$. It may
be noted that, except in phase-1, in all other phases the total magnetisation
vanishes.

The time dependent line magnetisation $m(y,t)={1 \over L}\Sigma_x s(x,y,t)$ 
is calculated and plotted against $y$ for different times. These are shown
in figure-2 for three different times ($t = 3000, 3030$ and 3060).
Here, $h_0=3.5$ and $w=1.0$. The propagating nature of the strips is clear from
the picture. To check the existence of propagating modes, for different sets of
values of $h_0$ and $w$, the line magnetisations 
($m(y,t)$) are plotted against $y$ for
different times. These are depicted in figure-3. Figure-3a and figure-3b shows
the variations of line magnetisation $m(y,t)$ with $y$ for different
times, $t=3000$ and $t=3030$ respectively. Here, $h_0=2.0$ and $w=3.0$.
This indicates the propagation. Figure-3c and figure-3d shows the same
(for different values of $h_0(=1.0$) and $w(=5.0$). This shows no propagation.

The four different dynamical phases are characterised by measuring the
structure factor $S(k,t) = {1 \over L}\int_0^L m(y,t)
 e^{iky} dy$. This integration is essentially a summation on the
lattice. It is noticed that
the magnitude of this structure factor, $|S(K,t)|$ takes different values
in different phases. This measurement is quit significant to get the
spin configurations in the lattice. If one calculate the  total magnetisation
and tries to use this to characterise the different dynamical phases, this
will not be successful. The total magnetisation will vanish for phase-2,
phase-3 and phase-4. The structure factor, $|S(k,t)|$ is a function of
$h_0$ and $w$. The phase-1 is characterise by $|S(k,t)|$=0. $|S(k,t)|$ will
assume a value above 0.5 in phase-2 and in phase-3 it will lie between
0.1 and 0.3. In phase-4, $|S(k,t)|$ will be very close to zero (but not
exactly equal to zero as was in phase-1). The structure factor was 
calculated after 6000 MCS and averaged over 10 different 
randomly disordered samples (for  
different random field configurations). 
The structure factor $|S(k,t)|$ is plotted, against
the width of the disorder (random field) for 
different values of $h_0$, in figure-4.
For $h_0=0.25$, the system remains in phase-1 
upto $w=3.7$ and then it transits to phase-4.
Now, for higher value of $h_0=1.75$, the system 
remains in phase-1 upto $w=2.1$, then it
enters into phase-3. Here the value of $|S(k,T)|$ becomes 
very close to 0.3. The system
remains in phase-3 upto $w=3.9$ and finally it 
transits to phase-4 (characterised by
$|S(k,t)| \simeq 0$). For this value of $h_0=1.75$, 
the phase-2 was not attained by the
system. Increasing the value of $h_0$ to 3.75, 
the variation of $|S(k,T)|$ with $w$, shows
that the system passes through all four 
nonequilibrium phases. In this case, transition from
phase-1 to phase-2 occurs at $w=0.3$, 
transition from phase-2 to phase-3 occurs at $w=3.9$
and finally the transition from phase-3 to phase-4 
occurs at $w=5.7$. A comprehensive
phase diagram is plotted and is shown in figure-5. 
This diagram shows the regions of four different phases 
in the $w-h_0$ plane.

The stability of the dynamical phases (mainly propagating, phase-2)
 in time is confirmed by a study for much longer
period ($t=30000$ MCS). The size dependence 
 was also studied for larger ($L=200$) lattices.
These are shown in figure-6.
This shows that the existence of spin wave propagating phase
 is neither a finite size 
effect (figure-6a) nor a transient phenomenon (figure-6b).

\vskip 1cm

\noindent {\bf IV. The summary:}

To summarize the results, the random (bimodal) field Ising
ferromagnet swept by propagating magnetic field wave is studied
by Monte Carlo simulations in two dimensions at zero temperature.
For a fixed value of the frequency and wavelegth of the propagating
magnetic field, the system resides in four distinct nonequilibrium
phases, depending upon the values of amplitude of the propagating magnetic
field and the width of the quenched random field. These four phases are:
phase-1, strongly correlated phase where all spins are parallel (up), 
phase-2, coherent spin wave (with sharp strips formed
by down spins) propagation, phase-3, coherent spin wave
(the boundary of strips is not quit sharp) propagation and phase-4,
the disordered phase, where a homogeneous distribution of up and down
spins is observed. A phase diagram is drawn in the plane formed by
the width of random field and the amplitude of propagating field.
These four phases are not arising due to finite size of the system.
Even, these are not transient states. These study is new and hope will
show some interesting nonequilibrium effects. 

Here, in this paper, the preliminary results are reported. There are some
important works have to be done in this respect. They are (i) to study the
velocity of spin wave propagation as a function of width of disorder,
(ii) the existence of diverging length scale (if any) along the phase
boundary\cite{maprop} (iii) the detailed finite size analysis and the scaling
behaviour. It is also important to check whether this type of propagating 
phases exist in the case of a Gaussian
distribution of random field. This requires lot of computational task to
get the phase diagram for these two types of distribution of random fields.
However, it seems that the boundaries will be moved towards
the right (that means transition will happen for higher values of the strength
of the random field) in the phase diagram. Since, the Gaussian distribution
will provide the major contribution of the random field in the central
region the higher values of strength is required to destroy the ordering.

The single crystal of $Mn12-ac$ and $LiHo_{x}Y_{1-x}F_4$ are two physical examples
of RFIM. The nonequilibrium phases discussed here may be observed in $Mn12-ac$
experimentally
by relaxation experiments using conventional SQUID magnetometer\cite{expt1}. In
the sample $LiHo_{x}Y_{1-x}F_4$ the propagating magnetic field would act like
time dependent perturbation (for very small amplitudes) and measuring the 
intensity variations, of the spectral lines of the transitions,
 the nonequilibrium phases may be observed\cite{expt2}.

\vskip 1cm

\noindent {Acknowledgements:} Author would like to thank Deepak Dhar,
Per Arne Rikvold and Bosiljka Tadic for useful discussions.

\newpage
\setlength{\unitlength}{0.240900pt}
\ifx\plotpoint\undefined\newsavebox{\plotpoint}\fi
\sbox{\plotpoint}{\rule[-0.200pt]{0.400pt}{0.400pt}}%


\vskip 1cm
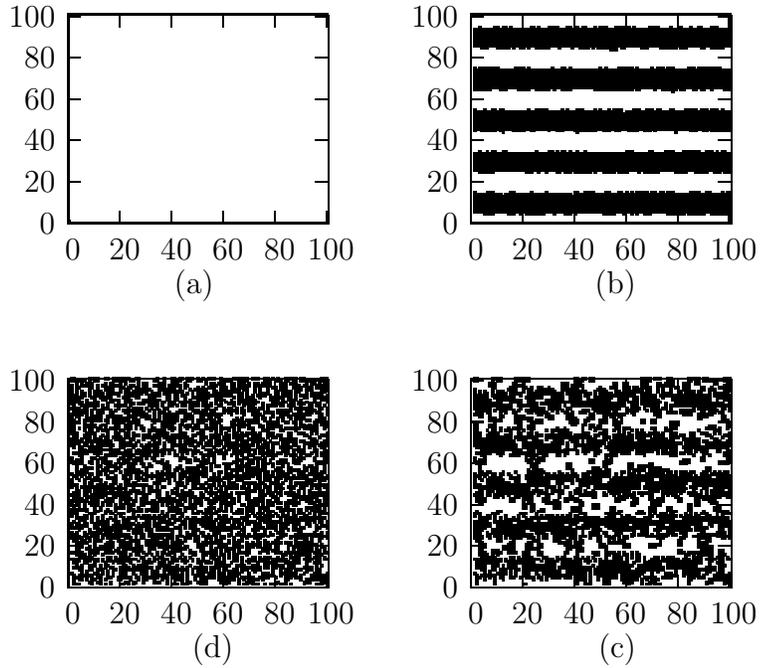
\noindent {\bf Fig-1.} Morphology of four different nonequilibrium
steady states. Here, $L=100$, $t=3000$ MCS, 
$f=0.01$, $\lambda=20$, $h_0=3.5$. Clockwise
from top-left, (a) phase of first kind (note that here all spins are
up and only the down spins are indicated by dots),
obtained for $w=0.4$, (b) phase of second
kind (propagating sharp strips) obtained for $w=1.0$, 
(c) phase of third kind (propagating strips of rough boundary)
 obtained for
$w=4.0$ and (d) phase of fourth kind, obtained for $w=6.0$.

\newpage

\setlength{\unitlength}{0.240900pt}
\ifx\plotpoint\undefined\newsavebox{\plotpoint}\fi
\sbox{\plotpoint}{\rule[-0.200pt]{0.400pt}{0.400pt}}%


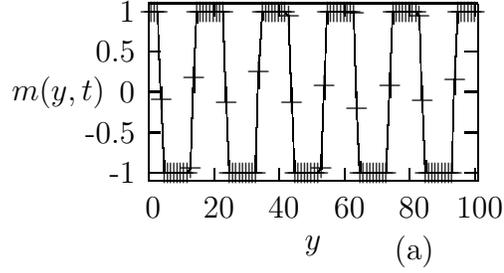
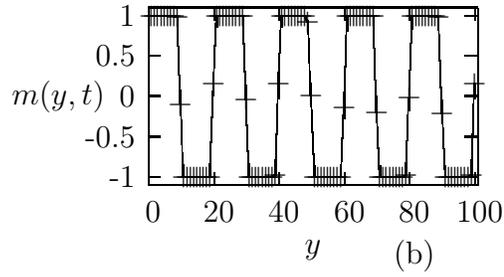
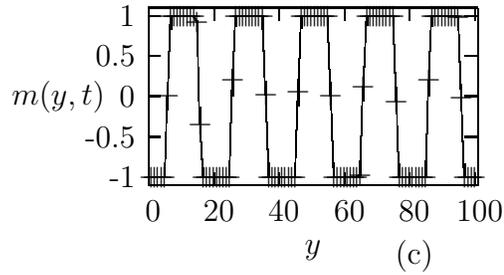
\noindent {\bf Fig-2.} The propagation of spin waves.
The value of $m(y,t)$ is plotted against $y$ for different
time $t$.
 Here,
$f=0.01$, $\lambda=20$, $h_0=3.5$, $w=1.0$. Different snapshots
are taken at different times (MCS). From top to bottom,
(a) for $t=3000$ MCS, (b) for $t=3030$ MCS and (c) for
$t=3060$ MCS.

\newpage
\setlength{\unitlength}{0.240900pt}
\ifx\plotpoint\undefined\newsavebox{\plotpoint}\fi
\sbox{\plotpoint}{\rule[-0.200pt]{0.400pt}{0.400pt}}%
%

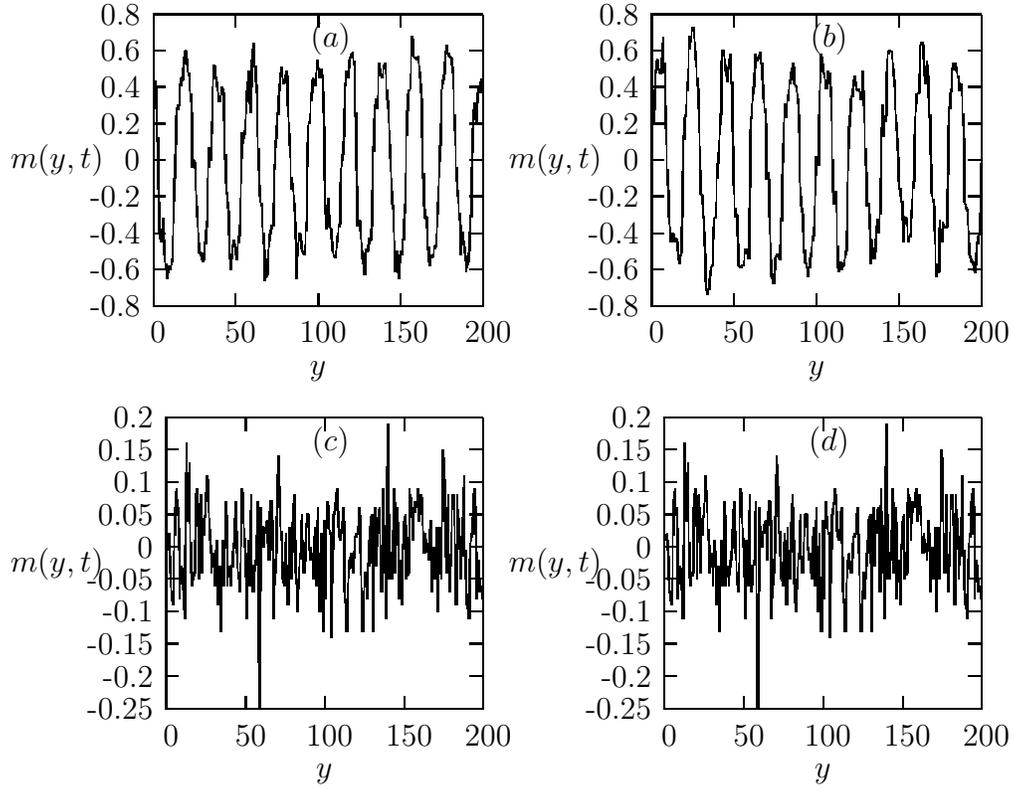
\noindent Fig-3. The line magnetisation $m(y,t)$ for different set of
values of parameters. (a) $h_0=2.0$, $w=3.0$ and $t=3000$ MCSS 
(b) $h_0=2.0$, $w=3.0$ and $t=3030$ MCSS, (c) $h_0=1.0$, $w=5.0$ and
$t=3000$ MCSS and (d) $h_0=1.0$, $w=5.0$ and $t=3030$ MCSS.

\newpage

\setlength{\unitlength}{0.240900pt}
\ifx\plotpoint\undefined\newsavebox{\plotpoint}\fi
\sbox{\plotpoint}{\rule[-0.200pt]{0.400pt}{0.400pt}}%
%

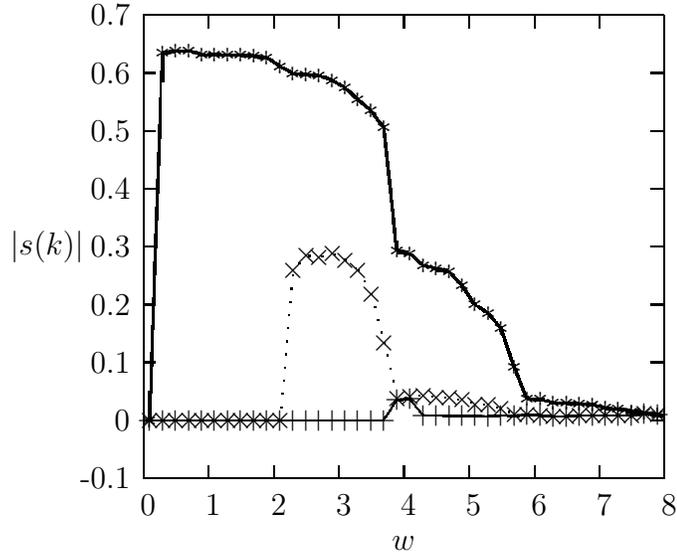
\noindent {\bf Fig-4.} The structure factor ($|s(k)|$) plotted against the
strength ($w$) of random field for different values of the amplitude
($h_0$) of the propagating magnetic field wave represented by different
symbols. Here, $h0=3.75 (\ast)$, $h_0=1.75 (\times)$ and $h_0=0.25 (+)$.
Continuous or dotted lines are guides to the eye. Here, 
$L=100$, $f=0.01$, $\lambda=20$. 

\newpage

\setlength{\unitlength}{0.240900pt}
\ifx\plotpoint\undefined\newsavebox{\plotpoint}\fi
\sbox{\plotpoint}{\rule[-0.200pt]{0.400pt}{0.400pt}}%
%

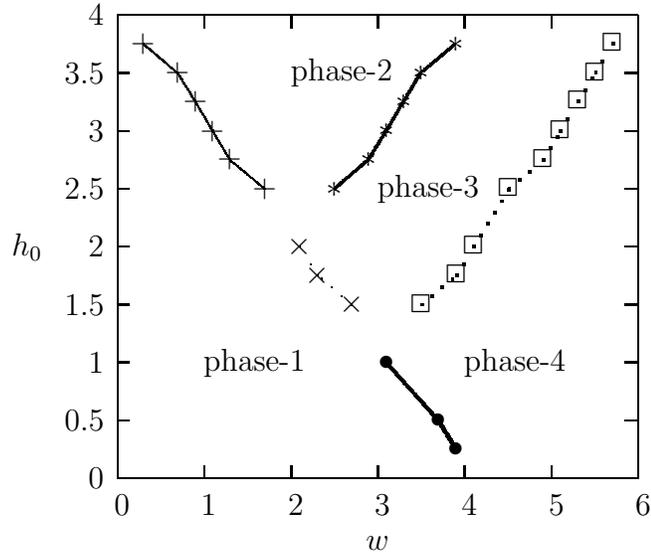
\noindent {\bf Fig-5.} The phase diagram in $w-h_0$ plane. Different
phase regions are indicated and the transition boundaries are shown by 
different symbols. Solid or dotted lines are guides to the eye.
Here, $L=100$, $f=0.01$, $\lambda=20$.

\newpage

\setlength{\unitlength}{0.240900pt}
\ifx\plotpoint\undefined\newsavebox{\plotpoint}\fi
\sbox{\plotpoint}{\rule[-0.200pt]{0.400pt}{0.400pt}}%
%

\vskip 0.5cm

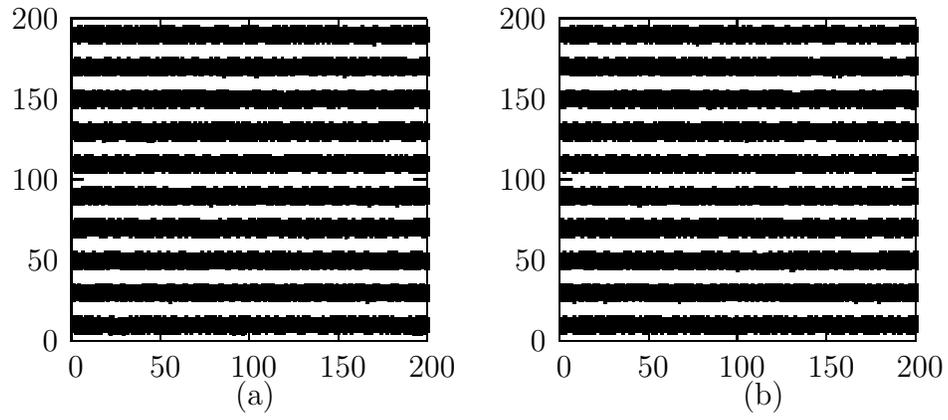
\noindent {Fig-6.} The steady state behaviour for different times. 
From top, (a) $t = 3000$ and
(b) $t=30000$. Here, $L=200$, $f=0.01$, $h_0=3.5$, $w=1.0$.
\end{document}